\documentclass[conference]{IEEEtran}

\renewcommand\IEEEkeywordsname{Index Terms}
\usepackage{graphicx}
\usepackage{subcaption}
\usepackage{tikz,xparse}

\usetikzlibrary{patterns}

\usetikzlibrary{dsp,chains}
\usetikzlibrary{matrix}
\usepackage{mathptmx}
\usepackage{verbatim}
\usepackage{calc}% http://ctan.org/pkg/calc
\usepackage{ifthen}
\usepackage{xifthen}
\usepackage{cancel}
\usepackage{bm}
\usepackage{verbatim}
\usepackage{multirow}
\usepackage{cite}
\usepackage[hyphens]{url}
\usepackage[nolist]{acronym} 
\usepackage{pgfplots}
\usetikzlibrary{arrows,shapes,graphs,graphs.standard,quotes,arrows.meta,decorations.markings,positioning}
\usetikzlibrary{intersections} % for named paths (arrows pointing at ellipse)
\usetikzlibrary{decorations, decorations.pathreplacing}

\usepackage{pifont}% http://ctan.org/pkg/pifont
\newcommand{\cmark}{\ding{51}}%
\newcommand{\xmark}{\ding{55}}%

\pgfplotsset{compat=newest}
\usepackage[bookmarks=false]{hyperref}
\usepackage{units}
\usepackage{amsmath, amsbsy, amssymb, latexsym }
\DeclareMathAlphabet{\mathcal}{OMS}{cmsy}{m}{n}
\hypersetup{bookmarksdepth=-2}
\usepackage{comment}
\usepackage[utf8]{inputenc}
\usepackage{xcolor}
\usepackage{enumitem}
\usepackage[linesnumbered,vlined]{algorithm2e}
\usepackage{algpseudocode}

\usepackage{marginnote}
\tikzset{>=latex}
\SetAlCapNameFnt{\footnotesize}
\SetAlCapFnt{\footnotesize}


\definecolor{mittelblau}{RGB}{0, 126, 198}
\definecolor{violettblau}{cmyk}{0.9, 0.6, 0, 0}
\definecolor{rot}{RGB}{238, 28 35}
\definecolor{apfelgruen}{RGB}{140, 198, 62}
\definecolor{gelb}{RGB}{255, 229, 0}
\definecolor{orange}{RGB}{244, 111, 33}
\definecolor{pink}{RGB}{237, 0, 140}
\definecolor{lila}{RGB}{128, 10, 145}
\definecolor{hellgrau}{RGB}{224, 224, 224}
\definecolor{mittelgrau}{RGB}{128, 128, 128}
\definecolor{dunkelgrau}{RGB}{80,80,80}
\definecolor{anthrazit}{RGB}{19, 31, 31}
\definecolor{darkgreen}{RGB}{34,139,34}
\definecolor{aqua}{RGB}{0, 255, 255}
\tikzset{
       vnd/.style={
        shape=circle,
        fill=black,
        draw,
        inner sep=0pt,
        minimum size=0.2cm},
        cnd/.style={
        shape=rectangle,
        fill=white,
        draw,
        minimum width=0.05mm,
        minimum height = 0.05mm}, 
         vndR/.style={
        shape=circle,
        fill=red,
        draw,
        inner sep=0pt,
        minimum size=0.2cm},
        cndR/.style={
        shape=rectangle,
        fill=white,
        draw=red,
        minimum width=0.05mm,
        minimum height = 0.05mm}
}

\newcommand{\qed}{\hfill\blacksquare}

\IEEEoverridecommandlockouts

\renewcommand{\vec}[1]{\mathbf{#1}}

\newcommand{\cv}{\vec{c}}

\newcommand{\uv}{\vec{u}}

\newcommand{\xv}{\vec{x}}
\newcommand{\yv}{\vec{y}}

\newcommand{\zerov}{\vec{0}}

\newcommand{\Gm}{\vec{G}}

\newcommand{\Id}{\vec{I}}

\newcommand{\Sm}{\vec{S}}
\newcommand{\Tm}{\vec{T}}

\newcommand{\LB}{\left(}
\newcommand{\RB}{\right)}

\renewcommand{\ln}[1]{\mathop{\mathrm{ln}}\LB #1\RB}

\begin{document}
	
\begin{NoHyper}
\title{Phase-Equivariant Polar Coded Modulation }

\author{\IEEEauthorblockN{Marvin Geiselhart, Marc Gauger, Felix Krieg, Jannis Clausius and Stephan ten Brink}
	\IEEEauthorblockA{
		Institute of Telecommunications, Pfaffenwaldring 47, University of  Stuttgart, 70569 Stuttgart, Germany 
		\\\{geiselhart,gauger,krieg,clausius,tenbrink\}@inue.uni-stuttgart.de\\
	}
    \thanks{This work is supported by the German Federal Ministry of Education and Research (BMBF) within the project Open6GHub (grant no. 16KISK019).}
}

\maketitle

\begin{acronym}
\acro{ML}{maximum likelihood}
\acro{BP}{belief propagation}
\acro{BPL}{belief propagation list}
\acro{LDPC}{low-density parity-check}
\acro{BER}{bit error rate}
\acro{SNR}{signal-to-noise-ratio}
\acro{BPSK}{binary phase shift keying}
\acro{AWGN}{additive white Gaussian noise}
\acro{LLR}{log-likelihood ratio}
\acro{MAP}{maximum a posteriori}
\acro{FER}{frame error rate}
\acro{BLER}{block error rate}
\acro{SCL}{successive cancellation list}
\acro{SC}{successive cancellation}
\acro{BI-DMC}{Binary Input Discrete Memoryless Channel}
\acro{CRC}{cyclic redundancy check}
\acro{CA-SCL}{CRC-aided successive cancellation list}
\acro{BEC}{Binary Erasure Channel}
\acro{BSC}{Binary Symmetric Channel}
\acro{BCH}{Bose-Chaudhuri-Hocquenghem}
\acro{RM}{Reed--Muller}
\acro{RS}{Reed-Solomon}
\acro{SISO}{soft-in/soft-out}
\acro{3GPP}{3rd Generation Partnership Project }
\acro{eMBB}{enhanced Mobile Broadband}
\acro{CN}{check node}
\acro{VN}{variable node}
\acro{GenAlg}{Genetic Algorithm}
\acro{CSI}{Channel State Information}
\acro{OSD}{ordered statistic decoding}
\acro{MWPC-BP}{minimum-weight parity-check BP}
\acro{FFG}{Forney-style factor graph}
\acro{MBBP}{multiple-bases belief propagation}
\acro{URLLC}{ultra-reliable low-latency communications}
\acro{mMTC}{massive machine-type communications}
\acro{DMC}{discrete memoryless channel}
\acro{SGD}{stochastic gradient descent}
\acro{QC}{quasi-cyclic}
\acro{NN}{neural network}
\acro{5G}{fifth generation mobile telecommunication}
\acro{SCAN}{soft cancellation}
\acro{AED}{automorphism ensemble decoding}
\acro{CCDF}{complementary cumulative distribution function}
\acro{VVPE}{Viterbi--Viterbi Phase Estimation}
\acro{QAM}{quadrature amplitude modulation}
\acro{QPSK}{quaternary phase shift keying}
\acro{PSK}{phase shift keying}
\acro{DPSK}{differential phase shift keying}
\acro{PAT}{pilot assisted transmission}
\acro{PM}{path metric}
\acro{RRC}{Rice--Rice--Cowley}
\acro{XOR}{exclusive or}
\end{acronym}

\begin{abstract}
For short-packet, low-latency communications over random access channels, piloting overhead significantly reduces spectral efficiency. Therefore, pilotless systems recently gained attraction.
While blind phase estimation algorithms such as \ac{VVPE} can correct a phase offset using only payload symbols, a phase ambiguity remains. 
We first show that the remaining phase rotations in a polar coded \ac{QAM} transmission with gray labeling are combinations of bit-flips and automorphisms. Therefore, the decoder is equivariant to such phase rotations and, by smartly selecting the frozen bits, one can jointly decode and resolve the phase ambiguity, without the need for pilot symbols or an outer code. Our proposed system outperforms pilot-assisted transmissions by up to 0.8\,dB and 2\,dB for \ac{QPSK} and 16-\ac{QAM}, respectively.

\end{abstract}
\acresetall

%\begin{IEEEkeywords}

%\end{IEEEkeywords}

\acresetall

\section{Introduction}
Recent trends in wireless networks, such as \ac{URLLC} and \ac{mMTC}, create an urgent need for the robust and efficient transmission of short packets. These scenarios typically involve a random access channel and due to the sporadic transmissions the receiver must re-estimate the channel for every transmission. Conventional systems rely on dedicated pilot symbols in the packet for this purpose, which is called \ac{PAT}. However, pilot symbols attribute a non-negligible overhead in the short-package systems, reducing the spectral efficiency.

Consequently, systems with smaller or no piloting overhead (also called ``non-coherent communication'') received increased attention.
A common approach is a two-stage estimation process.
The first stage is a \textit{blind} channel estimator that operates on all received symbols and is not aware which data symbols have been transmitted.
The second stage is the resolution of  the remaining phase ambiguity using a few pilot symbols or the error-correcting code.
For instance, in \cite{imad2010blind}, an \ac{LDPC} code is used to resolve the phase ambiguities. In \cite{PeihongNonCoherent}, the authors propose to use polar codes with an outer \ac{CRC} code for this purpose. Similarly, in \cite{ZhengHybridPolarCodingEstimation}, a polar codes were proposed for joint decoding and channel estimation of a fading channel.

In this work, we consider joint decoding and phase estimation by exploiting the symmetries of polar codes. In contrast to \cite{PeihongNonCoherent}, our approach poses little constraints on the polar code and does not require an outer code. Moreover, any polar decoder can be used, e.g.,
% ; for example,
a single low-complexity \ac{SC} decoder. However, a list decoder together with a \ac{CRC} can still be used to enhance the performance.

Polar code symmetries, namely their affine automorphism group  \cite{bardet_polar_automorphism, polar_aed, LiBLTA2021} have created interest due to applications in cryptoanalysis \cite{bardet_crypto} and low-latency permutation-aided decoder architectures \cite{polar_aed, pilletPolarCodesForAED}. Moreover, in \cite{polarNby4cyclic}, an automorphism corresponding to a quasi-cyclic shift by $N/4$ has been proposed to simplify the detection in a broadcast scenario.

\section{Preliminaries}\label{sec:preliminaries}
\subsection{System Model}
We consider a transmitter consisting of a linear channel encoder described by the generator matrix $\Gm$, encoding $K$ information bits $\uv$ into a codeword $\cv = \uv \Gm$ of length $N$. Assuming \ac{QAM}, information is encoded in both amplitude and phase of the transmit signal. The codeword is split into groups of $M_\mathrm{b}=\log_2(M)=\log_2(|\mathcal{K}|)$ bit, each mapped onto a complex symbol from the constellation $\mathcal{K}$. Fig.\,\ref{fig:constellations} shows examples of constellations for \ac{QPSK} and 16-\ac{QAM} as considered in this work. The mapping between bits $\mathbf{c}$ and symbols $\mathbf{x} = \operatorname{map}(\mathbf{c})$ corresponds to gray labeling.\footnote{For the proposed application in Section \ref{sec:main}, the gray labeling as in Fig. \ref{fig:constellations} must be used, i.e., without an interleaver.}
The signal $\xv$ is modulated to passband, transmitted over the channel, and at the receiver demodulated to the complex baseband again. An incorrectly assumed initial carrier phase at the receiver leads to a rotation of the constellation symbols in the complex plane. The equivalent baseband channel is modelled as
\begin{equation}
    y_i = x_i \cdot \mathrm{e}^{j\phi} + n_i 
\end{equation}
where $\phi \sim \mathcal{U}(0,2\pi)$ is a random uniform phase offset and $n_i \sim \mathcal{CN}(0,2\sigma^2)$ is complex \ac{AWGN} with noise power $N_0 = 2\sigma^2$.

The estimation and correction of the phase offset $\phi$ is the task of the receiver and is covered in Section \ref{ssec:blind} and \ref{sec:main}. The demapper obtains \acp{LLR} for each coded bit from the phase corrected received vector $\yv'$.
Under the assumption of equally likely transmission symbols, the \ac{MAP} rule for the  $k$-th bit of a symbol is
\begin{equation}
    \ell\left(y_i \right) \triangleq \ell\left( c_{i,k} \vert y_i \right) = \ln{\frac{\sum_{\hat{x}\in\mathcal{K}_{k,1}}p(y_i\vert x=\hat{x})}{\sum_{\hat{x}\in\mathcal{K}_{k,0}}p(y_i\vert x=\hat{x})}},
\end{equation}
where $\mathcal{K}_{k,0}$ and $\mathcal{K}_{k,1}$ denote the subsets of constellation points with the $k$-th bit equal to $0$ and $1$, respectively.
\begin{figure}[ht]
    \centering %
    \begin{subfigure}[t]{0.3046025\linewidth}
        \centering
        \resizebox{\linewidth}{!}{\begin{tikzpicture}[
singlearrow/.style = {-{Latex[length=2mm,width=2mm]}}
]
\def\labeling{{"00","01","10","11"}};
\newcommand{\getlabeling}[1]{\pgfmathparse{\labeling[#1]}\pgfmathresult};

\node (v4) at (0,2) {};
\node (v3) at (0,-2) {};
\node (v2) at (2,0) {};
\node (v1) at (-2,0) {};

\node (v4_pfusch) at (0,4*0.65/0.6) {};
\node (v3_pfusch) at (0,-4.2165*0.65/0.6) {};

\draw [singlearrow] (v1) -- (v2) node[at end, below,yshift=-0.1cm]{Re};
\draw [singlearrow] (v3) -- (v4) node[at end, right,xshift=0.1cm]{Im};

\phantom{
\draw [outer sep=0, line width= 0.65pt] (-2.3,-2.3*2*0.65/0.6) rectangle (2.3,2.3*2*0.65/0.6);
}
\foreach[count=\i] \y in {-1, 1} {
    \foreach[count=\j] \x in {-1, 1} {
	 \node[circle, radius=1pt, fill=black, label=above: \getlabeling{(\i-1)*2+(\j-1)}] 
at (\x,\y) {};		
    }
}

\end{tikzpicture} }
        \caption{\footnotesize QPSK}
        \label{fig:qpsk}
    \end{subfigure}
    \hfill
    \begin{subfigure}[t]{0.65\linewidth}
        \centering
        \resizebox{\linewidth}{!}{\begin{tikzpicture}[
singlearrow/.style = {-{Latex[length=2mm,width=2mm]}}
]
\def\labeling{{"0000","0001","0101","0100","0010","0011","0111","0110",
"1010","1011","1111","1110","1000","1001","1101","1100"}};	%gray

\newcommand{\getlabeling}[1]{\pgfmathparse{\labeling[#1]}\pgfmathresult};

\node (v4) at (0,4) {};
\node (v3) at (0,-4) {};
\node (v2) at (4,0) {};
\node (v1) at (-4,0) {};

\draw [singlearrow] (v1) -- (v2) node[at end, below,yshift=-0.1cm]{Re};
\draw [singlearrow] (v3) -- (v4) node[at end, right,xshift=0.1cm]{Im};

\foreach[count=\i] \y in {-3, -1, 1, 3} {
    \foreach[count=\j] \x in {-3, -1, 1, 3} {
	 \node[circle, radius=1pt, fill=black, label=above: \getlabeling{(\i-1)*4+(\j-1)}] 
at (\x,\y) {};		
    }
}
\phantom{
\draw [outer sep=0, line width=0.3pt] (-2.3*2*0.6/0.65,-2.3*2*0.6/0.65) rectangle (2.3*2*0.6/0.65,2.3*2*0.6/0.65);
}
\end{tikzpicture} }
        \caption{\footnotesize 16-QAM}
        \label{fig:16qam}
    \end{subfigure}
    \caption{\footnotesize Constellation diagrams with gray labeling.}
    \label{fig:constellations}
\end{figure}
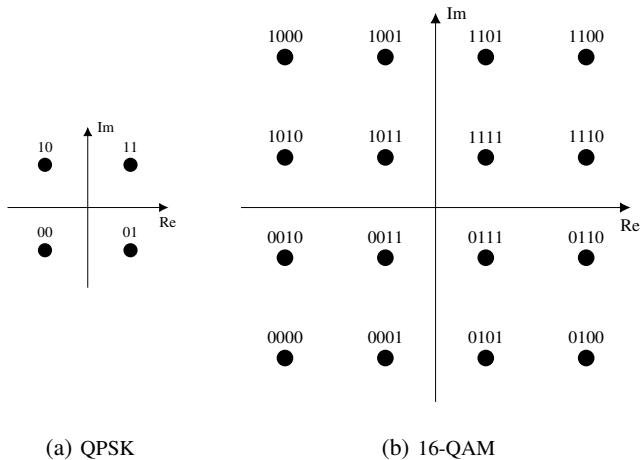

\subsection{Polar Codes}
Polar codes are based on the principle of channel polarization \cite{ArikanMain}. For a polar code of length $N=2^n$, the Plotkin transformation is applied $n$ times to obtain $N$ polarized \textit{synthetic} channels. $K$ polarized ``reliable'' channels are designated as the information set $\mathcal{I}$ carrying the information, while the remaining $N-K$ ``unreliable'' channels are frozen bits $\mathcal{F}$ and fixed to 0. Hence, the generator matrix is $\Gm$ are those rows of $\Gm_N = \left[\begin{smallmatrix} 1 & 0 \\ 1 & 1 \end{smallmatrix}\right]^{\otimes n}$ corresponding to $\mathcal{I}$, where $(\cdot)^{\otimes n}$ denotes the $n$-th Kronecker power.
A partial order on the reliability of the synthetic channels induces the affine automorphism group of the code \cite{bardet_polar_automorphism, bioglio2023groupproperties}, i.e., the set of codeword bit permutations that map the code onto itself.
A common decoder for polar codes is the \ac{SC} decoder \cite{ArikanMain}, where the information bits are decoded sequentially in natural order. An extension is \ac{SCL} decoding that keeps a list of the $L$ most promising decoding paths \cite{talvardylist}.

\subsection{Blind Phase Estimation}\label{ssec:blind}
\subsubsection{Maximum Likelihood}
Assuming $N_\mathrm{c}$ independent, uniformly distributed transmitted symbols, the log-likelihood function \cite{Georghiades1997BlindPE} of the phase angle $\phi$ is
\begin{equation}
    \ell(\phi) = \sum_{i=0}^{N_\mathrm{c}-1} \ln{ \sum_{x\in\mathcal{K}} \exp{\left(-\frac{1}{2\sigma^2}{\vert y_i - x \mathrm{e}^{j\phi} \vert}^2 \right)} }.
\end{equation}
Consequently, the \ac{ML} estimate of $\phi$ is
\begin{equation}
    \hat{\phi} = \arg\max_{\phi \in [0,2\pi)} \ell(\phi).
\end{equation}

If the constellation exhibits a rotational symmetry, the estimation can only be unique up to this symmetry.
In case of the fourfold-symmetric \ac{QPSK} and 16-\ac{QAM}, the phase decomposes into $\phi = \phi_\mathrm{i} + \phi_\mathrm{f}$ with the integer part $\phi_\mathrm{i}=m\frac{\pi}{2},\,m \in \{0,1,2,3\}$ and the fractional (fine) phase shift $\phi_\mathrm{f} \in [0,\nicefrac{\pi}{2})$. Only $\phi_\mathrm{f}$ can be estimated blindly, while $\phi_\mathrm{i}$ has to be recovered by different methods.

\subsubsection{Viterbi--Viterbi Algorithm}
In \cite{VVPE}, the calculation of a phase offset of an $M$-\acs{PSK} was presented.
The \ac{VVPE} takes advantage of the inherent symmetry of the \acs{PSK} constellation.
Independent of the transmitted symbols, multiplying the phase angle by $M$ rotates all symbols onto the same point, directly yielding an estimate of $M\phi$.
The blind fine estimation can be calculated to be
\begin{equation}
    \hat{\phi}_\mathrm{f} = \frac{1}{M}\operatorname{arctan}\left(
    \frac{
        \sum_{i=0}^{N_\mathrm{c}-1}\operatorname{Im}\left(\frac{y_i^{M}}{\lvert y_i \rvert^{M-1}}\right)
        }{
        % \frac{1}{N_\mathrm{c}}\cdot
        \sum_{i=0}^{N_\mathrm{c}-1}\operatorname{Re}\left(\frac{y_i^{M}}{\lvert y_i \rvert^{M-1}}\right)
        }
    \right).
\end{equation}
However, multiplying the phase by $M$, along with the final operation of dividing the $\arctan$ function by $M$, gives rise to an $M$-fold ambiguity in the phase estimate \cite{VVPE}.

\subsubsection{Rice--Rice--Cowley Algorithm}
To estimate the phase offset of higher \ac{QAM} modulations, the blind estimation scheme needs to be adapted.
In \cite{RiceRiceCowleyBlindEstimation16QAM}, the authors treat an 16-\ac{QAM} as a combination of three individual \acp{QPSK} in order to perform the phase estimation.
We denote this estimation scheme as \ac{RRC} estimation. In a first step, the received symbols are classified into the three rings by their amplitude.
The phases of the eight symbols on the innermost and outermost ring are jointly estimated, as they stem from the same (scaled) \ac{QPSK} constellation with the same phase offset.
Using this first estimate, the remaining eight symbols on the middle circle are partitioned into two sub-constellations whose offsets are again estimated by \ac{VVPE}. With this additional information, the initial phase estimation can be improved.

\section{Phase-Equivariant Polar Coded Modulation}\label{sec:main}

\subsection{Symmetries of Polar-Coded Modulation}

\textbf{Lemma 1 (2-Cyclic Polar Codes):}
Let $\mathcal{C}$ be a polar code with information set $\mathcal{I}$ such that $i | 1 \in \mathcal{I} \; \forall i \in \mathcal{I}$, where ``$|$'' denotes the \textit{bitwise or} operation. Then for all $\cv \in \mathcal{C}$, also $\cv' = [c_{(i+1)\operatorname{mod} 2}]_i \in \mathcal{C}$. In other words, if for each even information bit index $i$ also $i+1$ corresponds to an information bit, then the code is quasi-cyclic with block-size 2, i.e., the permutation
\begin{equation}
    \tau = (1,0,3,2,5,4,\dots,N-1,N-2)
\end{equation}
is an automorphism of $\mathcal{C}$. We call such a polar code \textit{2-cyclic}. Note that all polar codes constructed for \ac{DMC} exhibit this property, as the synthetic channel $i+1$ is always more reliable than $i$, for even $i$ \cite{ArikanMain}.

\textit{Proof.} The statement is a relaxed version of \cite[Thm. 2]{bardet_polar_automorphism} for weakly decreasing codes. Intuitively, $\tau$ affects solely the very first stage of the polar factor graph, so only the $N=2$ component codes have to be regarded. The swap of the two coded bits is an automorphism, if each component code is either rate-0, rate-1 or a repetition code, i.e., $\mathcal{I}_{N-2} \in \{ \varnothing, \{1\}, \{0,1\}\}$.$\qed$

Furthermore, we know that an automorphism on the codeword bits (represented by the $N\times N$ permutation matrix $\Tm$) corresponds to a linear transformation $\Sm$ of the message bits \cite[Ch. 8, §5, Lemma 12]{macwilliams77}
\begin{equation}
    \uv \cdot \Gm = \cv \quad \Longleftrightarrow \quad  \uv \cdot \Sm \cdot \Gm = \cv \cdot \Tm. \label{eq:scramble}
\end{equation}
For polar codes and the 2-cyclic shift $\tau$, we find this transformation matrix as \cite{PolarsubcodeBPL}
\begin{equation}
    \tilde{\Sm} = \Gm_N \Tm \Gm_N = \left[\begin{matrix} 1 & 1 \\ 0 & 1 \end{matrix}\right] \otimes \Id_{N/2} . \label{eq:polarscramble}
\end{equation}
Note that in this case $\tilde{\Sm}$ is also of dimension $N \times N$ and includes also the frozen bits. $\Sm$ is obtained by selecting only the rows and columns corresponding to $\mathcal{I}$.

\textbf{Theorem 1 (Rotational Symmetry):} Let $\mathcal{C}$ be a 2-cyclic polar code with $N\ge4$ and information set $\mathcal{I}$. Furthermore, let $\mathcal{X} = \{\xv=\operatorname{map}(\cv),  \cv \in \mathcal{C}\}$ be the set of all complex transmit signals under one of the mappings in Fig.\,\ref{fig:constellations} and let $\mathcal{R}=\{N-2,N-1\}$ for \ac{QPSK} and $\mathcal{R}=\{N-4,N-3\}$ for 16-\ac{QAM}. Then
\begin{equation}
    \xv \mathrm{e}^{-j\frac{\pi}{2}} \in \mathcal{X} \quad \forall \xv \in \mathcal{X},
\end{equation}
In other words, a mapped codeword is transformed to another mapped codeword under the rotation $\mathrm{e}^{-j\frac{\pi}{2}}$.

\textit{Proof.}
For \ac{QPSK}, the rotation of $\pi/2$ corresponds to swapping adjacent bits and then flipping every other bit, i.e.,
\begin{equation}
    (c_0,c_1,c_2,c_3,\dots) \mapsto (\bar{c}_1,c_0,\bar{c}_3,c_2,\dots).
\end{equation}
This is a combination of the addition of row $N-2$ of $\Gm_N$ and the 2-cyclic automorphism $\tau$ from Lemma 1. From the requirements of Lemma 1, also $N-1$ must be in $\mathcal{I}$.
For 16-\ac{QAM} with the labeling from Fig.\,\ref{fig:16qam}, the $\pi/2$-rotation implies
\begin{equation}
    (c_0,c_1,c_2,c_3,\dots) \mapsto (\bar{c}_1,c_0,c_3,c_2,\dots),
\end{equation}
corresponding to $\tau$ from Lemma 1 followed by the addition of row $N-4$ of $\Gm_N$. Again, then also $N-3$ must be in $\mathcal{I}$. $\qed$

Remark: A similar statement holds for the rotation by $\pi$ in case of the \ac{BPSK}. Here, observe that this corresponds to flipping all bits, i.e., the addition of the all-one codeword. This codeword is the last row of $\Gm_N$ and thus part of the code if $N-1 \in \mathcal{I}$. 

\subsection{Joint Decoding and Phase Ambiguity Resolution}\label{ssec:phase_resolve}
\begin{figure}
    \centering
    \resizebox{\linewidth}{!}{\begin{tikzpicture}[
	demapper/.pic = {
		\begin{axis}[xshift=-0.4cm, yshift=-0.2cm,
		width=0.8cm, height=0.8cm,
		scale only axis, hide axis, clip=false,
		xmin=-1.2, xmax=1.2, ymin=-1.2, ymax=1.2,
		domain=-1:1, samples=100]
		\addplot [color=black, mark=*] coordinates {(-0.707,0.707)};
		\addplot [color=black, mark=*] coordinates {(-0.707,-0.707)};
		\addplot [color=black, mark=*] coordinates {(0.707,0.707)};
		\addplot [color=black, mark=*] coordinates {(0.707,-0.707)};
		\addplot[thin] coordinates {(0,-1.2) (0,1.2)};
		\addplot[thin] coordinates {(-1.2,0) (1.2,0)};
		\node at (axis cs:0,-2) {\footnotesize Demapper};
		\end{axis}
	},
	block/.style={draw, rectangle, minimum size=1.4cm, align=center, execute at begin node={\baselineskip=1em}},
	doublearrow/.style = {-{Latex[length=2mm,width=2mm]},double},
	doubleline/.style = {-,double},
	singlearrow/.style = {-{Latex[length=1.5mm,width=1.5mm]}}
	]
	
	\node (y) at (0.8,0) {$\yv$};
	\node[dspnodefull](s1) at (1.3,0) {};
	\node[dspmixer](m1) at (4,0) {};
	\node[block](FPE) at (2.5,-2) {\footnotesize Blind\\\footnotesize Fine Phase\\\footnotesize Estimation};
	\node[draw](exp) at (4,-1) {\footnotesize $\mathrm{e}^{-j\hat{\phi}_\mathrm{f}}$};
	\node[dspnodefull, label=below:$\hat{\phi}_\mathrm{f}$](s2) at (4,-2) {};
	\node[block](demapper)at (5.6,0){}; \pic at (demapper){demapper};
	\node[block](decoder)at (8,0){\footnotesize Polar\\\footnotesize Decoder}; 
	\node[dspnodefull](s3) at (9,-0.2) {};
	\node[block](flip)at (10.5,0){$\Sm^{\hat{u}'_{\mathrm{R},0}} $};
	\node[block](compphi) at (10.5,-2){\footnotesize Compute \\\footnotesize $\hat{\phi}$}; 
	
	\node (uhat) at (12,0) {$\hat{\uv}$};
	\node (phihat) at (12,-2) {$\hat{\phi}$};
	
	\draw[doubleline] (y) -- (s1);
	\draw[doublearrow] (s1) |- (FPE);
	\draw[doublearrow] (s1) -- (m1);
	\draw[doublearrow] (m1) --node[above]  {$\yv'$} (demapper);
	\draw[doublearrow] (exp) -- (m1);
	\draw (FPE) -- (s2);
	\draw[singlearrow] (s2) -- (exp);
	\draw[singlearrow] (s2) -- (compphi);
	\draw[singlearrow] (compphi) -- (phihat);
	\draw[singlearrow] (demapper) --node[above]  {$\ell(\yv')$} (decoder);
    \draw[singlearrow] ([yshift=0.2 cm]decoder.east) --node[above]  {$\hat{\uv}'$} ([yshift=0.2 cm]flip.west);
    \draw ([yshift=-0.2 cm]decoder.east) -- (s3);
    \draw[singlearrow](s3) -- ([yshift=-0.2 cm]flip.west);
    \draw[singlearrow](s3) |- node[right,yshift=1cm] {$\hat{\uv}'_\mathrm{R}$} ([yshift=0.3 cm]compphi.west);

    \draw[singlearrow] (flip) -- (uhat);
	\draw[singlearrow] (compphi) -- (phihat);
	
\end{tikzpicture}}
    \caption{\footnotesize Block Diagram of joint decoding and phase estimation at the receiver.}
    \label{fig:block_diagram}
\end{figure}
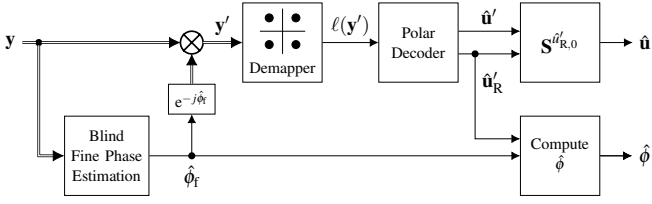
Based on Theorem 1, we propose a method for pilotless transmission that resolves the phase ambiguities of the blind phase estimator after polar decoding. We again let $\mathcal{R}=\{N-2,N-1\}$ for \ac{QPSK} and $\mathcal{R}=\{N-4,N-3\}$ for 16-\ac{QAM} and define a vector of rotation discriminating bit channels $\uv_\mathrm{R}$ that allows a recovery of the phase. In particular, $\uv_\mathrm{R,QPSK}=(u_{N-2},u_{N-1})$ and $\uv_\mathrm{R,16-QAM}=(u_{N-4},u_{N-3})$ for \ac{QPSK} and 16-\ac{QAM}, respectively.
Notice that if at the transmitter $\uv_\mathrm{R}=\zerov$, a phase rotation by multiples of $\pi/2$ results in a nonzero estimate $\hat{\uv}_\mathrm{R}$ at the receiver, namely $\hat{\uv}_\mathrm{R}$ indicates the number of $\pi/2$-rotations in binary. This property directly follows from Theorem 1, as a single rotation flips bit $\uv_{\mathrm{R},0}$ and a double rotation flips bit $\uv_{\mathrm{R},1}$. However, due to the 2-cyclic automorphism $\tau$, also other bits of the estimate $\hat{\uv}$ are affected, which must be undone by applying the matrix $\Sm$ from Eq. \ref{eq:polarscramble} on $\hat{\uv}$ (or, $\tilde{\Sm}$, respectively on the full $\hat{\uv}$). Our proposed joint estimation and decoding scheme can thus be summarized as follows:
\begin{enumerate}
    \item Design a 2-cyclic polar code $\mathcal{C}$, where $\mathcal{R} \subseteq \mathcal{F}$. A practical approach is to use a code design with $K+2$ information bits and freeze the bits $\mathcal{R}$ afterwards.
    \item At the transmitter, encode using $\mathcal{C}$, followed by either \ac{QPSK} or 16-QAM mapping according to Fig. \ref{fig:constellations}, i.e., $\xv = \operatorname{map}(\uv \Gm)$.
    \item At the receiver, perform blind fine phase estimation on the received symbols $\yv$, e.g., using \ac{ML} or \ac{VVPE} and correct the fine phase offset $\hat{\phi}_\mathrm{f}$ to obtain $\yv'$ which is aligned to the constellation up to rotations of $\frac{m\pi}{2}$.
    \item Demap and decode $\yv'$ using any polar decoder using information set $\mathcal{I}'=\mathcal{I} \cup \mathcal{R}$. The information bit estimates corresponding to $\mathcal{I}$ are denoted by $\hat{\uv}'$, and the phase recovery bits corresponding to $\mathcal{R}$ by $\hat{\uv}'_\mathrm{R}$.
    \item Recover the remaining phase rotation as $\hat{\phi}_\mathrm{i} = \frac{m\pi}{2}$ with $m=2\hat{u}'_{\mathrm{R},1}+\hat{u}'_{\mathrm{R},0}$. The total phase rotation is $\hat{\phi} = \hat{\phi}_\mathrm{i}+\hat{\phi}_\mathrm{f}$.
    \item Resolve the phase ambiguity in the message: $\hat{\uv} = \hat{\uv}' \cdot \Sm^{m}$. Note that $\Sm^2 = \Id$ and, thus, $\Sm^{m}=\Sm^{\hat{u}'_{\mathrm{R},0}}$, which can be implemented using at most $K/2$ binary additions.
\end{enumerate}
Fig. \ref{fig:block_diagram} depicts a block diagram of the proposed receiver.

Remark: Again, the same idea can be applied to \ac{BPSK} modulation. In this case, $\mathcal{R}=\{N-1\}$ and only phase ambiguities of $\pi$ have to be considered. As no permutation of the coded bits happens, no correction matrix $\Sm$ has to be applied, futher simplifying the system.

\subsection{List and Ensemble Decoding}
\begin{figure}
    \centering
    \resizebox{\linewidth}{!}{\begin{tikzpicture}[
	demapper/.pic = {
		\begin{axis}[xshift=-0.4cm, yshift=-0.2cm,
		width=0.8cm, height=0.8cm,
		scale only axis, hide axis, clip=false,
		xmin=-1.2, xmax=1.2, ymin=-1.2, ymax=1.2,
		domain=-1:1, samples=100]
		\addplot [color=black, mark=*] coordinates {(-0.707,0.707)};
		\addplot [color=black, mark=*] coordinates {(-0.707,-0.707)};
		\addplot [color=black, mark=*] coordinates {(0.707,0.707)};
		\addplot [color=black, mark=*] coordinates {(0.707,-0.707)};
		\addplot[thin] coordinates {(0,-1.2) (0,1.2)};
		\addplot[thin] coordinates {(-1.2,0) (1.2,0)};
		\node at (axis cs:0,-2) {\footnotesize Demapper};
		\end{axis}
	},
	block/.style={draw, rectangle, minimum size=1.4cm, align=center, execute at begin node={\baselineskip=1em}},
	block_small/.style={draw, rectangle, minimum size = 1cm, align=center, execute at begin node={\baselineskip=1em}},
	doublearrow/.style = {-{Latex[length=2mm,width=2mm]},double},
	doubleline/.style = {-,double},
	singlearrow/.style = {-{Latex[length=1.5mm,width=1.5mm]}},
	mydiamond/.style={draw, diamond, aspect=2.5,text width=1.5cm, inner sep=0pt},
	]

	\node (y) at (2.5,-2.5) {$\yv$};
	\node[dspnodefull](s1) at (3,-2.5) {};
	\draw[doubleline] (y) --  (s1);
	
	\node[block, minimum height=7cm](decoder)at (7.8,-2.3){\footnotesize Polar\\ \footnotesize List \\ \footnotesize / \\\footnotesize  Ensemble \\\footnotesize  Decoder}; 
	
	\node[block_small, minimum height=7cm](ml) at (14,-2.3){\footnotesize min\\ \footnotesize PM};

	\node (exp0) at (3.7,1) {$\mathrm{e}^{j0}$};
	\node[dspmixer](m0) at (3.7,0) {};
	\node[block](demapper0)at (5.3,0){}; \pic at (demapper0){demapper};
	\node[block_small](flip0) at (9.9,0){\footnotesize $\Sm^{\hat{u}'_{\mathrm{R},0,0}} $};
	\node[mydiamond](check0) at (12,0){\footnotesize CRC check};
	\node(fail0) at (12,-0.8) {};
	
	\draw[doublearrow] (m0) --node[above]  {$\yv'_0$} (demapper0);
	\draw[doublearrow] (exp0) -- (m0);
	\draw[singlearrow] (demapper0) --node[above]  {$\ell(\yv'_0)$} (decoder.west |- demapper0);
    \draw[singlearrow] ([yshift=0.2 cm]decoder.east |- flip0.west) --node[above]  {$\hat{\uv}'_0$} ([yshift=0.2 cm]flip0.west);
    \draw[singlearrow] ([yshift=-0.2 cm]decoder.east |- flip0.west) --node[below]  {$\hat{\uv}'_{\mathrm{R},0}$} ([yshift=-0.2 cm]flip0.west);
    \draw[singlearrow] (flip0) --node[above]  {$\hat{\uv}_0$} (check0);
    \draw[singlearrow] (check0) --node[above] {\cmark} (ml.west |- check0);
    \draw[singlearrow] (check0) --node[right] {\xmark} (fail0);
	
	\node (exp1) at (3.7,-1) {$\mathrm{e}^{-j\frac{\pi}{2L}}$};
	\node[dspmixer](m1) at (3.7,-2) {};
	\node[block](demapper1)at (5.3,-2){}; \pic at (demapper1){demapper};
	\node[block_small](flip1) at (9.9,-2){\footnotesize $\Sm^{\hat{u}'_{\mathrm{R},1,0}} $};
	\node[mydiamond](check1) at (12,-2){\footnotesize CRC check};
	\node(fail1) at (12,-2.8) {};
	
	\draw[doublearrow] (m1) --node[above]  {$\yv'_1$} (demapper1);
	\draw[doublearrow] (exp1) -- (m1);
	\draw[singlearrow] (demapper1) --node[above]  {$\ell(\yv'_1)$} (decoder.west |- demapper1);
    \draw[singlearrow] ([yshift=0.2 cm]decoder.east |- flip1.west) --node[above]  {$\hat{\uv}'_1$} ([yshift=0.2 cm]flip1.west);
    \draw[singlearrow] ([yshift=-0.2 cm]decoder.east |- flip1.west) --node[below]  {$\hat{\uv}'_{\mathrm{R},1}$} ([yshift=-0.2 cm]flip1.west);
     \draw[singlearrow] (flip1) --node[above]  {$\hat{\uv}_1$} (check1);
    \draw[singlearrow] (check1) --node[above] {\cmark} (ml.west |- check1);
    \draw[singlearrow] (check1) --node[right] {\xmark} (fail1);
	
	\node (exp2) at (3.7,-4) {$\mathrm{e}^{-j\frac{(L-1)\pi}{2L}}$};
	\node[dspmixer](m2) at (3.7,-5) {};
	\node[block](demapper2)at (5.3,-5){}; \pic at (demapper2){demapper};
	\node[block_small](flip2) at (9.9,-5){\footnotesize $\Sm^{\hat{u}'_{\mathrm{R},L-1,0}} $};
	\node[mydiamond](check2) at (12,-5){\footnotesize CRC check};
	\node(fail2) at (12,-5.8) {};
	
	\draw[doublearrow] (m2) --node[above]  {$\yv'_{L-1}$} (demapper2);
	\draw[doublearrow] (exp2) -- (m2);
	\draw[singlearrow] (demapper2) --node[above]  {$\ell(\yv'_{L-1})$} (decoder.west |- demapper2);
    \draw[singlearrow] ([yshift=0.2 cm]decoder.east |- flip2.west) --node[above]  {$\hat{\uv}'_{L-1}$} ([yshift=0.2 cm]flip2.west);
    \draw[singlearrow] ([yshift=-0.2 cm] decoder.east |- flip2.west) --node[below,xshift=0.1cm]  {$\hat{\uv}'_{\mathrm{R},L-1}$} ( [yshift=-0.2 cm] flip2.west);
	\draw[singlearrow] (flip2) --node[above, xshift=.1cm]  {$\hat{\uv}_{L-1}$} (check2);
    \draw[singlearrow] (check2) --node[above] {\cmark} (ml.west |- check2);
	\draw[singlearrow] (check2) --node[right] {\xmark} (fail2);

	\draw[doublearrow] (s1) |- (m0);
	\draw[doublearrow] (s1) |- (m1);
	\draw[doublearrow] (s1) |- (m2);

    \draw[singlearrow] ([yshift=.8cm] decoder.east |- flip0) --node[above]  {$\mathrm{PM}_0$} ([yshift=0.8cm] ml.west |- flip0);
    \draw[singlearrow] ([yshift=.8cm] decoder.east |- flip1) --node[above]  {$\mathrm{PM}_1$} ([yshift=0.8cm] ml.west |- flip1);
    \draw[singlearrow] ([yshift=.8cm] decoder.east |- flip2) --node[above]  {$\mathrm{PM}_{L-1}$} ([yshift=0.8cm] ml.west |- flip2);
    
    \node() at (5.3,-3.5) {$\vdots$};
    \node() at (9.7,-3.5) {$\vdots$};
    \node() at (12,-3.5) {$\vdots$};

    \node (uhat) at (15,-2.5) {$\hat{\uv}$};
    \draw[singlearrow] (ml.east |- uhat) -- (uhat);
	
\end{tikzpicture}}
    \caption{\footnotesize Modified receiver structure for list or ensemble decoding without blind phase estimation.}
    \label{fig:block_diagram_list}
\end{figure}
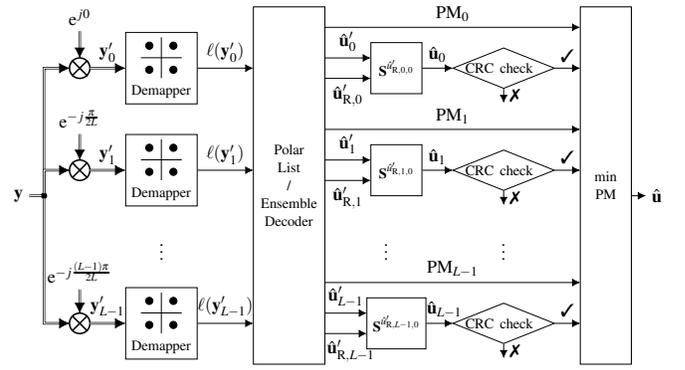
The symmetry from Theorem 1 is a polar code property and thus, the proposed scheme works with any polar decoder. However, when an outer code is required in the polar decoding, such as \ac{CRC}-aided \ac{SCL}, the system has to be modified. In general, the permutation involved in the $\pi/2$ rotation is not an automorphism of the concatenated code and, hence,  $\hat{\uv}'$ is not a codeword of the outer code. For this reason, the outer code membership must be checked after resolving the phase ambiguity for each candidate in the list.

With this addition, the error-rate performance of the system (up to the 2 bit rate loss for $\uv_\mathrm{R}$) is only limited by the accuracy of the blind fine phase estimator. For short blocklengths, this performance limitation may not be desired. Therefore, we propose a second system compatible with \ac{SCL} decoding without blind phase estimation. 

Fig. \ref{fig:block_diagram_list} shows the block diagram of the modified receiver using list decoding. While the standard \ac{SCL} decoder uses only a single input, variants have been proposed that initialize all $L$ paths with different $\yv$, generated from codeword permutations \cite{permuteRM, SCAL}. The decoding paths from the different inputs compete with each other whenever the decoding paths branch and only the $L$ most promising paths based on a \ac{PM} are kept. In this work, we initialize the $L$ paths with the different hypotheses for $\phi$, ranging from $0$ to $\pi/2$. At the decoder output, the message candidate $\hat{\uv}_l$ with the smallest final \ac{PM} is declared as the final decoding result. For this message, the remaining phase ambiguity is resolved as presented in Section \ref{ssec:phase_resolve}. When an outer \ac{CRC} code is used, the ambiguity is resolved for each candidate before removing the candidates failing the \ac{CRC} check and \ac{PM}-based selection.

Another decoding option is an ensemble decoder, i.e., a collection of low-complexity independent (e.g., \ac{SC}) decoders \cite{rm_automorphism_ensemble_decoding}. Compared to \ac{SCL} decoding, all decoding paths are separate and no branching is performed, reducing the complexity and decoding latency at the cost of a slight error-rate performance degradation at the same list size \cite{SCAL}.

\section{Results}

\subsection{Baseline System}
The error-rate performance of the proposed system is evaluated using polar codes according to the 5G design \cite{polar5G2018}.
As reference, we compare to multiple possible baseline systems:
\begin{enumerate}
    \item A scenario where $\phi$ is perfectly known at the receiver, corresponding to an \ac{AWGN} channel only.
    \item A \ac{PAT} system without any blind estimation, i.e., $\phi$ is only estimated from $p$ pilot systems. To make the comparison fair, the polar code has been shortened by $p\cdot M_\mathrm{b}$ bit such that this system has the same overall rate. The number of pilots $p$ is optimized for best \ac{BLER} performance in each case. This case is denoted by ``$p$-PAT''.
    \item A hybrid \ac{PAT} system with blind estimation, i.e., $\phi_\mathrm{f}$ is estimated using a blind estimator and the remaining phase ambiguity is resolved using $p$ pilots. Like in the previous system, rate-matching has been optimized. This case is denoted by ``$p$-PAT+ML/VVPE/RRC'', depending on the blind estimation algorithm used.
\end{enumerate}

\subsection{Error Rate Performance}
\begin{figure}[ht]
    \centering
    \resizebox{\linewidth}{!}{\begin{tikzpicture}
\begin{axis}[
	width=\linewidth,
	height=.6\linewidth,
	grid style={dotted,gray},
	xmajorgrids,
	yminorticks=true,
	ymajorgrids,
	legend columns=1,
	legend pos=south west,   
	legend cell align={left},
	xlabel={$E_\mathrm{s}/N_0$ in dB},
	ylabel={BLER},
	legend image post style={mark indices={}},
	legend style={fill opacity=0.7, text opacity=1},
	ymode=log,
	mark size=2pt,
	mark options={solid},
	xmin=1,
	xmax=6.3,
	ymin=1e-04,
	ymax=1
]

\addplot [color=red,line width = 1pt,dashed,mark=triangle]
table[col sep=comma]{
0.00, 9.751e-01
0.50, 9.516e-01
1.00, 8.750e-01
1.50, 7.802e-01
2.00, 6.133e-01
2.50, 4.238e-01
3.00, 2.478e-01
3.50, 1.182e-01
4.00, 5.253e-02
4.50, 1.709e-02
5.00, 5.062e-03
5.50, 1.176e-03
6.00, 2.526e-04
6.50, 4.613e-05
};
\label{plot:QPSK_SC_10_Pilots}
\addlegendentry{\footnotesize 10-\ac{PAT}};

\addplot [color=apfelgruen,line width = 1pt,solid,mark=square]
table[col sep=comma]{
0.00, 9.039e-01
0.50, 8.589e-01
1.00, 6.887e-01
1.50, 5.254e-01
2.00, 3.458e-01
2.50, 2.360e-01
3.00, 1.141e-01
3.50, 5.203e-02
4.00, 1.938e-02
4.50, 5.786e-03
5.00, 1.285e-03
5.50, 2.811e-04
6.00, 5.207e-05
};
\label{plot:QPSK_SC_VVPE_5_Pilots}
\addlegendentry{\footnotesize 5-\ac{PAT}+\ac{VVPE}}

\addplot [color=mittelblau,line width = 1pt,solid,mark=o]
table[col sep=comma]{
0.00, 8.688e-01
0.50, 7.692e-01
1.00, 6.286e-01
1.50, 4.438e-01
2.00, 3.011e-01
2.50, 1.575e-01
3.00, 6.711e-02
3.50, 2.690e-02
4.00, 7.898e-03
4.50, 2.099e-03
5.00, 4.231e-04
5.50, 7.085e-05
};
\label{plot:QPSK_SC_EV}
\addlegendentry{\footnotesize Joint+VVPE}

\addplot [color=black,line width = 1pt,solid]
table[col sep=comma]{
0.00, 7.365e-01
0.50, 6.026e-01
1.00, 4.099e-01
1.50, 2.686e-01
2.00, 1.370e-01
2.50, 6.110e-02
3.00, 2.424e-02
3.50, 7.490e-03
4.00, 1.921e-03
4.50, 4.835e-04
5.00, 9.569e-05
};
\label{plot:QPSK_SC_Genie}
\addlegendentry{\footnotesize AWGN only}

\draw[-latex, line width=1pt, color=black] (axis cs: 5.38, 1.7e-3) --node[right,yshift=0cm, xshift=0.6cm, color=black]  {\footnotesize $0.8$\,dB} (axis cs: 4.58, 1.7e-3);

\draw[-latex, line width=1pt, color=black] (axis cs: 5.2, 7e-4) --node[right,yshift=0cm, xshift=0.2cm, color=black]  {\footnotesize $0.3$\,dB} (axis cs: 4.85, 7e-4);

\end{axis}
\end{tikzpicture}}
    \caption{\footnotesize \ac{BLER} performance comparison for \ac{QPSK} based transmission of $K=64$ information bits over $N_\mathrm{c}=64$ complex channel uses (code rate $R=\nicefrac{1}{2}$, \ac{SC} decoding).}
    \label{fig:bler_qpsk}
\end{figure}
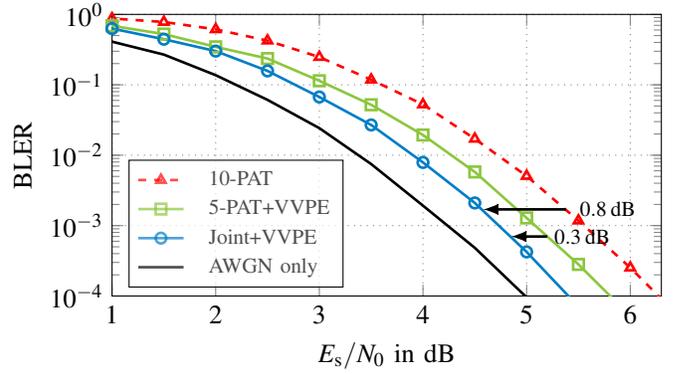

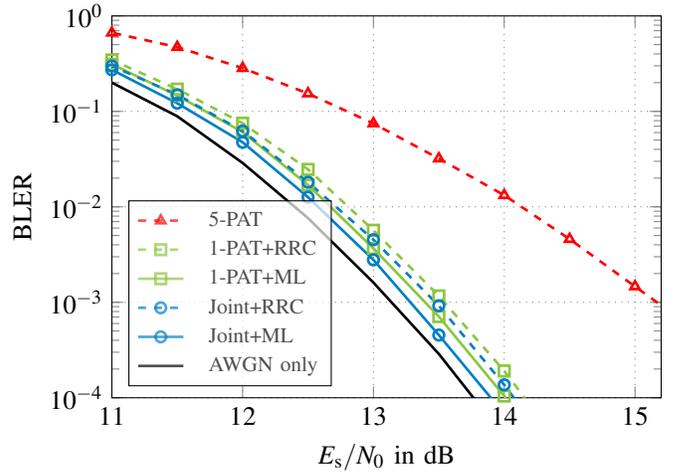
\begin{figure}[ht]
    \centering
    \resizebox{\linewidth}{!}{\begin{tikzpicture}
\begin{axis}[
	width=\linewidth,
	height=.75\linewidth,
	grid style={dotted,gray},
	xmajorgrids,
	yminorticks=true,
	ymajorgrids,
	legend columns=1,
	legend pos=south west,   
	legend cell align={left},
	legend style={fill opacity=0.7, text opacity=1},
	xlabel={$E_\mathrm{s}/N_0$ in dB},
	ylabel={BLER},
	legend image post style={mark indices={}},
	ymode=log,
	mark size=2pt,
	mark options={solid},
	xmin=11,
	xmax=15.2,
	ymin=1e-04,
	ymax=1
]

\addplot [color=red,line width = 1pt,dashed,mark=triangle]
table[col sep=comma]{
7.00, 1.000e+00
7.50, 1.000e+00
8.00, 1.000e+00
8.50, 9.980e-01
9.00, 9.961e-01
9.50, 9.808e-01
10.00, 9.324e-01
10.50, 8.361e-01
11.00, 6.662e-01
11.50, 4.715e-01
12.00, 2.839e-01
12.50, 1.539e-01
13.00, 7.474e-02
13.50, 3.205e-02
14.00, 1.323e-02
14.50, 4.592e-03
15.00, 1.465e-03
15.50, 4.699e-04
16.00, 1.271e-04
16.50, 3.069e-05
};
\label{plot:116QAM_SC_5_Pilots}
\addlegendentry{\footnotesize 5-\ac{PAT}};

\addplot [color=apfelgruen,line width = 1pt,dashed,mark=square]
table[col sep=comma]{
7.00, 1.000e+00
7.50, 1.000e+00
8.00, 1.000e+00
8.50, 9.980e-01
9.00, 9.658e-01
9.50, 9.107e-01
10.00, 7.779e-01
10.50, 5.657e-01
11.00, 3.459e-01
11.50, 1.710e-01
12.00, 7.516e-02
12.50, 2.466e-02
13.00, 5.680e-03
13.50, 1.167e-03
14.00, 1.908e-04
14.50, 2.393e-05
};
\label{plot:16QAM_SC_VVPE_1_Pilots}
\addlegendentry{\footnotesize 1-\ac{PAT}+RRC}

\addplot [color=apfelgruen,line width = 1pt,solid,mark=square]
table[col sep=comma]{
7.00, 1.000e+00
7.50, 1.000e+00
8.00, 9.980e-01
8.50, 9.942e-01
9.00, 9.678e-01
9.50, 8.839e-01
10.00, 7.282e-01
10.50, 5.288e-01
11.00, 3.117e-01
11.50, 1.463e-01
12.00, 6.062e-02
12.50, 1.690e-02
13.00, 3.672e-03
13.50, 7.087e-04
14.00, 1.042e-04
14.50, 1.146e-05
};
\label{plot:16QAM_SC_ML_1_Pilots}
\addlegendentry{\footnotesize 1-\ac{PAT}+\ac{ML}}

\addplot [color=mittelblau,line width = 1pt,dashed,mark=o]
table[col sep=comma]{
7.00, 1.000e+00
7.50, 1.000e+00
8.00, 9.980e-01
8.50, 9.846e-01
9.00, 9.678e-01
9.50, 8.791e-01
10.00, 7.175e-01
10.50, 5.282e-01
11.00, 2.989e-01
11.50, 1.497e-01
12.00, 6.231e-02
12.50, 1.812e-02
13.00, 4.526e-03
13.50, 9.203e-04
14.00, 1.366e-04
14.50, 1.625e-05
};
\label{plot:16QAM_SC_EV_RRC}
\addlegendentry{\footnotesize Joint+RRC}

\addplot [color=mittelblau,line width = 1pt,solid,mark=o]
table[col sep=comma]{
7.00, 1.000e+00
7.50, 9.980e-01
8.00, 9.980e-01
8.50, 9.903e-01
9.00, 9.587e-01
9.50, 8.729e-01
10.00, 7.011e-01
10.50, 4.946e-01
11.00, 2.725e-01
11.50, 1.218e-01
12.00, 4.746e-02
12.50, 1.261e-02
13.00, 2.784e-03
13.50, 4.538e-04
14.00, 6.764e-05
};
\label{plot:16QAM_SC_EV_ML}
\addlegendentry{\footnotesize Joint+ML}

\addplot [color=black,line width = 1pt,solid]
table[col sep=comma]{
7.00, 1.000e+00
7.50, 9.980e-01
8.00, 9.931e-01
8.50, 9.702e-01
9.00, 9.215e-01
9.50, 7.843e-01
10.00, 5.877e-01
10.50, 3.681e-01
11.00, 2.001e-01
11.50, 8.889e-02
12.00, 2.880e-02
12.50, 7.614e-03
13.00, 1.609e-03
13.50, 2.877e-04
14.00, 3.894e-05
};
\label{plot:16QAM_SC_Genie}
\addlegendentry{\footnotesize AWGN only}

\end{axis}
\end{tikzpicture}}
    \caption{\footnotesize \ac{BLER} performance comparison for 16-\ac{QAM} based transmission of $K=192$ information bits over $N_\mathrm{c}=64$ complex channel uses (code rate $R=\nicefrac{3}{4}$, \ac{SC} decoding).}
    \label{fig:bler_16qam}
\end{figure}

Fig. \ref{fig:bler_qpsk} shows the \ac{BLER} performance of \ac{PAT} and the proposed joint estimation and decoding for \ac{QPSK} transmission of $K=64$ bits over $N_\mathrm{c}=64$ channel uses. All systems use \ac{SC} decoding. The baseline \ac{PAT} system requires $p=10$ pilot symbols for best performance, while blind \ac{VVPE} gains $0.5$\,dB by reducing the required pilots to 5. Our proposed pilotless equivariant system outperforms this hybrid \ac{PAT} by another $0.3$\,dB, with a remaining gap to the case without phase offset of another $0.5$\,dB at \ac{BLER} of $10^{-3}$.
In Fig. \ref{fig:bler_16qam} the same setup is used with 16-\ac{QAM} to achieve a target spectral efficiency of $3$ bit per channel use (bpcu) ($K=192$, $N_\mathrm{c}=64$). Both \ac{RRC} and \ac{ML} blind phase estimation are considered. The proposed joint system with \ac{ML} blind phase estimation performs within $0.15$\,dB of the \ac{AWGN} case. Moreover, the joint method outperforms the hybrid \ac{PAT} by $0.1$\,dB; in both cases, \ac{RRC} estimation loses roughly $0.15$\,dB versus the \ac{ML} estimator. Here, the optimal \ac{PAT} system would require $p=5$ pilots and $2$\,dB higher \ac{SNR} for a  \ac{BLER} of $10^{-3}$.

\subsection{List Decoding}
\begin{figure}
    \centering
    \resizebox{\linewidth}{!}{\begin{tikzpicture}
\begin{axis}[
	width=\linewidth,
	height=.75\linewidth,
	grid style={dotted,gray},
	xmajorgrids,
	yminorticks=true,
	ymajorgrids,
	legend columns=1,
	legend pos=south west,   
	legend cell align={left},
	xlabel={$E_\mathrm{s}/N_0$ in dB},
	ylabel={BLER},
	legend image post style={mark indices={}},
	ymode=log,
	mark size=2pt,
	mark options={solid},
	xmin=8,
	xmax=15,
	ymin=1e-04,
	ymax=4e-1,
	xtick={8,9,10,11,12,13,14,15},
]

\addplot [color=red,line width = 1pt,dashed,mark=triangle]
table[col sep=comma]{
5.00, 9.789e-01
5.50, 9.204e-01
6.00, 8.481e-01
6.50, 7.533e-01
7.00, 6.259e-01
7.50, 4.638e-01
8.00, 3.329e-01
8.50, 2.523e-01
9.00, 1.548e-01
9.50, 8.729e-02
10.00, 4.844e-02
10.50, 2.364e-02
11.00, 9.946e-03
11.50, 4.475e-03
12.00, 1.735e-03
12.50, 5.970e-04
13.00, 1.844e-04
13.50, 6.609e-05
};
\label{plot:116QAM_SCL_5_Pilots}

\addplot [color=red,line width = 1pt,solid,mark=triangle]
table[col sep=comma]{
5.00, 9.788e-01
5.50, 9.375e-01
6.00, 8.671e-01
6.50, 7.991e-01
7.00, 7.131e-01
7.50, 5.751e-01
8.00, 4.251e-01
8.50, 2.767e-01
9.00, 1.672e-01
9.50, 8.271e-02
10.00, 4.194e-02
10.50, 1.852e-02
11.00, 7.028e-03
11.50, 2.781e-03
12.00, 9.518e-04
12.50, 3.408e-04
13.00, 9.278e-05
};
\label{plot:116QAM_CA_SCL_5_Pilots}

\addplot [color=apfelgruen,line width=1pt,dashed,mark=+,mark phase=1,mark repeat=2]
table[col sep=comma]{
5.00, 8.613e-01
5.50, 7.790e-01
6.00, 6.964e-01
6.50, 5.895e-01
7.00, 4.777e-01
7.50, 3.851e-01
8.00, 2.754e-01
8.50, 1.992e-01
9.00, 1.398e-01
9.50, 9.109e-02
10.00, 6.050e-02
10.50, 3.751e-02
11.00, 2.165e-02
11.50, 1.296e-02
12.00, 7.016e-03
12.50, 3.730e-03
13.00, 1.904e-03
13.50, 1.009e-03
14.00, 5.010e-04
14.50, 2.249e-04
15.00, 9.728e-05
};
\label{plot:16QAM_SCL_RRC_1_Pilots}

\addplot [color=apfelgruen,line width=1pt,solid,mark=x,mark phase=2,mark repeat=2]
table[col sep=comma]{
5.00, 8.613e-01
5.50, 7.790e-01
6.00, 6.964e-01
6.50, 5.895e-01
7.00, 4.777e-01
7.50, 3.851e-01
8.00, 2.754e-01
8.50, 1.992e-01
9.00, 1.398e-01
9.50, 9.109e-02
10.00, 6.050e-02
10.50, 3.751e-02
11.00, 2.165e-02
11.50, 1.296e-02
12.00, 7.016e-03
12.50, 3.730e-03
13.00, 1.904e-03
13.50, 1.009e-03
14.00, 5.010e-04
14.50, 2.249e-04
15.00, 9.728e-05
};
\label{plot:16QAM_CA_SCL_RRC_1_Pilots}

\addplot [color=apfelgruen,line width=1pt,dashed,mark=square,mark phase=1,mark repeat=2]
table[col sep=comma]{
5.00, 8.528e-01
5.50, 7.582e-01
6.00, 6.810e-01
6.50, 5.730e-01
7.00, 4.507e-01
7.50, 3.445e-01
8.00, 2.391e-01
8.50, 1.714e-01
9.00, 1.156e-01
9.50, 7.074e-02
10.00, 4.634e-02
10.50, 2.712e-02
11.00, 1.573e-02
11.50, 8.311e-03
12.00, 4.320e-03
12.50, 1.854e-03
13.00, 6.688e-04
13.50, 1.911e-04
14.00, 4.821e-05
};
\label{plot:16QAM_SCL_ML_1_Pilots}

\addplot [color=apfelgruen,line width=1pt,solid,mark=diamond,mark phase=1,mark repeat=2]
table[col sep=comma]{
5.00, 9.073e-01
5.50, 8.322e-01
6.00, 7.434e-01
6.50, 6.201e-01
7.00, 4.800e-01
7.50, 3.394e-01
8.00, 2.257e-01
8.50, 1.384e-01
9.00, 9.060e-02
9.50, 6.016e-02
10.00, 3.928e-02
10.50, 2.483e-02
11.00, 1.562e-02
11.50, 8.284e-03
12.00, 3.942e-03
12.50, 1.719e-03
13.00, 6.458e-04
13.50, 1.887e-04
14.00, 5.051e-05
};
\label{plot:16QAM_CA_SCL_ML_1_Pilots}

\addplot [color=magenta,line width=1pt,dashed,mark=*]
table[col sep=comma]{
5.00, 8.372e-01
5.50, 7.631e-01
6.00, 6.680e-01
6.50, 5.504e-01
7.00, 3.918e-01
7.50, 2.724e-01
8.00, 1.609e-01
8.50, 9.089e-02
9.00, 4.791e-02
9.50, 2.190e-02
10.00, 9.609e-03
10.50, 3.628e-03
11.00, 1.212e-03
11.50, 3.820e-04
12.00, 1.221e-04
12.50, 3.493e-05
};
\label{plot:16QAM_SCE_EV}

\addplot [color=magenta,line width=1pt,solid,mark=*]
table[col sep=comma]{
5.00, 9.427e-01
5.50, 8.912e-01
6.00, 8.128e-01
6.50, 7.145e-01
7.00, 5.372e-01
7.50, 3.903e-01
8.00, 2.509e-01
8.50, 1.469e-01
9.00, 7.693e-02
9.50, 3.473e-02
10.00, 1.355e-02
10.50, 4.425e-03
11.00, 1.332e-03
11.50, 3.131e-04
12.00, 8.216e-05
};
\label{plot:16QAM_CA_SCE_EV}

\addplot [color=mittelblau,line width=1pt,dashed,mark=o]
table[col sep=comma]{
5.00, 8.446e-01
5.50, 7.427e-01
6.00, 6.151e-01
6.50, 4.898e-01
7.00, 3.376e-01
7.50, 2.061e-01
8.00, 1.187e-01
8.50, 7.121e-02
9.00, 3.617e-02
9.50, 1.689e-02
10.00, 7.416e-03
10.50, 2.743e-03
11.00, 1.073e-03
11.50, 3.281e-04
12.00, 1.132e-04
12.50, 3.295e-05
};
\label{plot:16QAM_SCL_EV}

\addplot [color=mittelblau,line width=1pt,solid,mark=o]
table[col sep=comma]{
5.00, 8.934e-01
5.50, 8.073e-01
6.00, 6.942e-01
6.50, 5.277e-01
7.00, 3.682e-01
7.50, 2.281e-01
8.00, 1.275e-01
8.50, 5.827e-02
9.00, 2.182e-02
9.50, 6.631e-03
10.00, 1.731e-03
10.50, 3.919e-04
11.00, 7.774e-05
};
\label{plot:16QAM_CA_SCL_EV}

\addplot [color=black,line width=1pt,dashed]
table[col sep=comma]{
5.00, 7.800e-01
5.50, 6.454e-01
6.00, 5.045e-01
6.50, 3.476e-01
7.00, 2.481e-01
7.50, 1.589e-01
8.00, 9.089e-02
8.50, 5.098e-02
9.00, 2.467e-02
9.50, 1.086e-02
10.00, 4.416e-03
10.50, 1.461e-03
11.00, 4.951e-04
11.50, 1.404e-04
12.00, 3.594e-05
};
\label{plot:16QAM_SCL_Genie}

\addplot [color=black,line width=1pt,solid]
table[col sep=comma]{
5.00, 7.740e-01
5.50, 6.392e-01
6.00, 4.751e-01
6.50, 3.140e-01
7.00, 1.887e-01
7.50, 9.675e-02
8.00, 4.003e-02
8.50, 1.427e-02
9.00, 3.932e-03
9.50, 9.021e-04
10.00, 1.886e-04
10.50, 2.676e-05
};
\label{plot:16QAM_CASCL_Genie}

\coordinate (legend) at (axis description cs:0.98,0.98);

\draw[-latex, line width=1pt, color=black] (axis cs: 12.55, 3e-4) --node[above,yshift=0cm, xshift=.4cm, color=black]  {\footnotesize $2$\,dB} (axis cs: 10.6, 3e-4);

\draw[-latex, line width=1pt, color=black] (axis cs: 13, 2e-4) --node[below,yshift=0cm, xshift=.1cm, color=black]  {\footnotesize $1.3$\,dB} (axis cs: 11.7, 2e-4);

\end{axis}

\matrix [
draw,
fill=white,
fill opacity=0.7,
text opacity=1,
matrix of nodes,
align =left,
column sep = -5,
inner sep= 2,
anchor=north east,
font=\footnotesize,
column 1/.style={anchor=base west},
mark options={solid},
] at (legend) {
	& no CRC & CRC-7 \\
	5-PAT & \ref{plot:116QAM_SCL_5_Pilots} & \ref{plot:116QAM_CA_SCL_5_Pilots}\\
	1-PAT+RRC & \ref{plot:16QAM_SCL_RRC_1_Pilots} & \ref{plot:16QAM_CA_SCL_RRC_1_Pilots}\\
	1-PAT+ML & \ref{plot:16QAM_SCL_ML_1_Pilots} & \ref{plot:16QAM_CA_SCL_ML_1_Pilots}\\
	Joint, ens. & \ref{plot:16QAM_SCE_EV} & \ref{plot:16QAM_CA_SCE_EV}\\
	Joint, list & \ref{plot:16QAM_SCL_EV} & \ref{plot:16QAM_CA_SCL_EV}\\
	AWGN only & \ref{plot:16QAM_SCL_Genie} & \ref{plot:16QAM_CASCL_Genie}\\
};

\end{tikzpicture}}
    \caption{\footnotesize \ac{BLER} performance comparison for 16-\ac{QAM} based transmission of $K=64$ information bits over $N_\mathrm{c}=32$ complex channel uses (code rate $R=\nicefrac{1}{2}$, \ac{SC}-based list and ensemble decoding with $L=8$).}
    \label{fig:bler_16qam_list}
\end{figure}
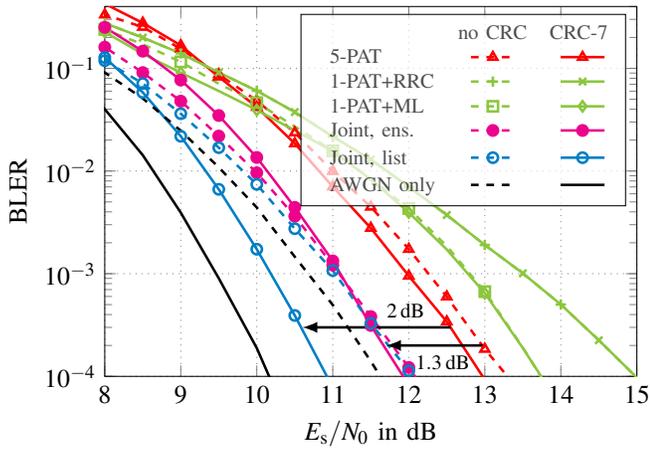

Fig. \ref{fig:bler_16qam_list} compares the performance of \ac{PAT} and joint estimation and decoding for the case of list and ensemble decoding with $L=8$. Systems with and without a 7-bit \ac{CRC} are examined; the \ac{CRC} generator polynomial is $g(z) = z^7 + z^3 +1$ \cite{liva2016}. 
The considered scenario is the transmission of $K=64$ information bits over $N_\mathrm{c}=128$ complex channel uses, i.e., a spectral efficiency of $2$\,bpcu.  In this scenario, the main source of error is in the phase estimation. Hence, the \ac{CRC} does not improve the performance of the hybrid \ac{PAT} systems and shows a small gain in case of standard \ac{PAT}, which outperforms the hybrid approach. The proposed joint estimation and decoding without any blind estimation however, outperforms \ac{PAT} by $1.3$\,dB and $2$\,dB, with and without \ac{CRC}, respectively. As expected, the ensemble decoder cannot profit from the outer \ac{CRC} code, in contrast to the list decoder. However, if no \ac{CRC} is used, the ensemble decoder performs very similar to the list decoder and should hence be preferred in this case due to the lower implementation complexity and latency.

\section{Conclusion}\label{sec:conc}
In this paper, we presented a joint estimation and decoding framework for channels with phase offset. The system works by exploiting phase-equivariant symmetries of polar-coded transmission with \ac{QPSK} and 16-\ac{QAM}. An extension for list and ensemble decoding has been proposed, significantly outperforming \ac{PAT} in \ac{BLER} performance. 

While this work considered only single-shot processing, extensions to iterative estimation and decoding (e.g., using soft-in/soft-out decoders) may also profit from the phase-equivariant properties. Additionally, future work may focus on other channels, such as block fading channels or multipath channels. Lastly, different outer codes can be considered that respect the 2-cyclic polar code automorphism (similar to \cite{GelincikPilletDynamicFrozen}) and thus, may be checked before resolving the phase ambiguities to reduce implementation complexity.

\bibliographystyle{IEEEtran}
\bibliography{references}
\end{NoHyper}
\end{document}